\documentclass{mem}
\usepackage{natbib}\usepackage{txfonts}\usepackage{balance}
\usepackage{graphicx}
\usepackage[a4paper]{hyperref}
\idline{1}{1}

\def\lesssim{\mathrel{\hbox{\rlap{\hbox{\lower4pt\hbox{$\sim$}}}\hbox{$<$}}}}
\def\gtrsim{\mathrel{\hbox{\rlap{\hbox{\lower4pt\hbox{$\sim$}}}\hbox{$>$}}}}
\def\LCDM{$\Lambda$CDM}
\def\ergscm{erg~s$^{-1}$~cm$^{-2}$}
\def\fun{{\rm erg\, s}^{-1}{\rm cm}^{-2}}
\def\arcsec{$^{\prime\prime}$}

\def\deg{$^o$}
\def\Chandra{{\it Chandra~}}
\def\XMM{{\it XMM-Newton~}}

\begin{document}

\title{
Wide Field X-ray Telescope: Mission Overview
}


\author{
P.\, Rosati\inst{1}\and S.\, Borgani\inst{2,3}\and R.\, Gilli\inst{4}\and M.\,Paolillo\inst{5}\and P.\, Tozzi\inst{6} \and S.\, Murray\inst{7} \and R.\, Giacconi\inst{7} \and A.\, Ptak\inst{7} \and M.\, Weisskopf\inst{8}, W.\, Forman\inst{9} \and C.\, Jones\inst{9}  \and the WFXT Team
          }


\institute{
       ESO-European Southern Observatory, D-85748 Garching bei
       M\"unchen, Germany 
  \and Dipartimento di Fisica, Sezione di Astronomia, Universit\`a
       di Trieste, Via Tiepolo 11, I-34143 Trieste, Italy
  \and INFN, Sezione di Trieste, Via Valerio 2, I-34127 Trieste, Italy
  \and INAF-Osservatorio Astronomico di Bologna, Via Ranzani 1,
      I-40127 Bologna, Italy
  \and Universit\`a Federico II, Dip. di Scienze Fisiche, Via Cintia, I-80126, 
       Napoli, Italy
 \and INAF-Osservatorio Astronomico di Trieste, Via Tiepolo 11,
      I-34143, Trieste, Italy
 \and Department of Physics and Astronomy, The Johns Hopkins University,
       Baltimore MD, USA
 \and NASA Marshall Space Flight Center, Space Science Office, VP62, 
      Huntsville, AL 35812, USA
 \and Harvard-Smithsonian Center for Astrophysics, Cambridge, MA, USA
} 
\authorrunning{Rosati et al.}

\titlerunning{Mission Overview}

\abstract{ The Wide Field X-Ray Telescope (WFXT) is a medium-class
  mission designed to be 2-orders-of-magnitude more sensitive than any
  previous or planned X-ray mission for large area surveys and to
  match in sensitivity the next generation of wide-area optical, IR
  and radio surveys. Using an innovative wide-field X-ray optics
  design, WFXT provides a field of view of 1 square degree (10 times
  Chandra) with an angular resolution of 5\arcsec (Half Energy Width,
  HEW) nearly constant over the entire field of view, and a large
  collecting area (up to 1 m$^2$ at 1 keV, $>10$x Chandra) over the 0.1-7 keV
  band. WFXT’s low-Earth orbit also minimizes the particle
  background. In five years of operation, WFXT will carry out three
  extragalactic surveys at unprecedented depth and address outstanding
  questions in astrophysics, cosmology and fundamental physics. In this
  article, we illustrate the mission concept and the connection
  between science requirements and mission parameters. }

\maketitle

\section{Mission concept}

Exploring the high-redshift Universe, to the epochs of cluster
formation all the way back to the primordial populations of galaxies
and super massive black holes (SMBHs) requires sensitive, high angular
resolution, wide X-ray surveys to complement deep, wide-field surveys
in other wavebands. The Wide Field X-Ray Telescope (WFXT) was designed
to be 2-orders-of-magnitude more sensitive than any previous or
planned X-ray mission for large area surveys and to match in
sensitivity the next generation of wide-area optical, IR and radio
surveys. In its current concept \citep{Murray08}
\footnote{http://www.wfxt.eu, \, http://wfxt.pha.jhu.edu}, WFXT is a
medium-class PI mission with a broad science grasp which will build a
unique astrophysical data set, consisting of $\gtrsim 5 \times 10^5$
clusters of galaxies to $z\sim2$, $>10^7$ AGN to $z>6$, and $\sim10^5$
normal and starburst galaxies at $z\lesssim 1$.  These large samples will
provide a description of the cosmic evolution of baryons, map the
large scale structure of the Universe, constrain and test cosmological
models and fundamental physics (such as the nature of Dark Matter,
Dark Energy and gravity), determine the black hole accretion history
to early epochs and its intimate link with galaxy formation, and
provide an unprecedented view of nearby galaxies including our
own. The science breadth of WFXT is only outlined below and fully described
by the specific contributions in this volume, which span a range of
prominent science cases. 

\begin{figure*}[t!]
\begin{center}
\raisebox{2.4cm}{\includegraphics[width=0.51\textwidth,clip=true]{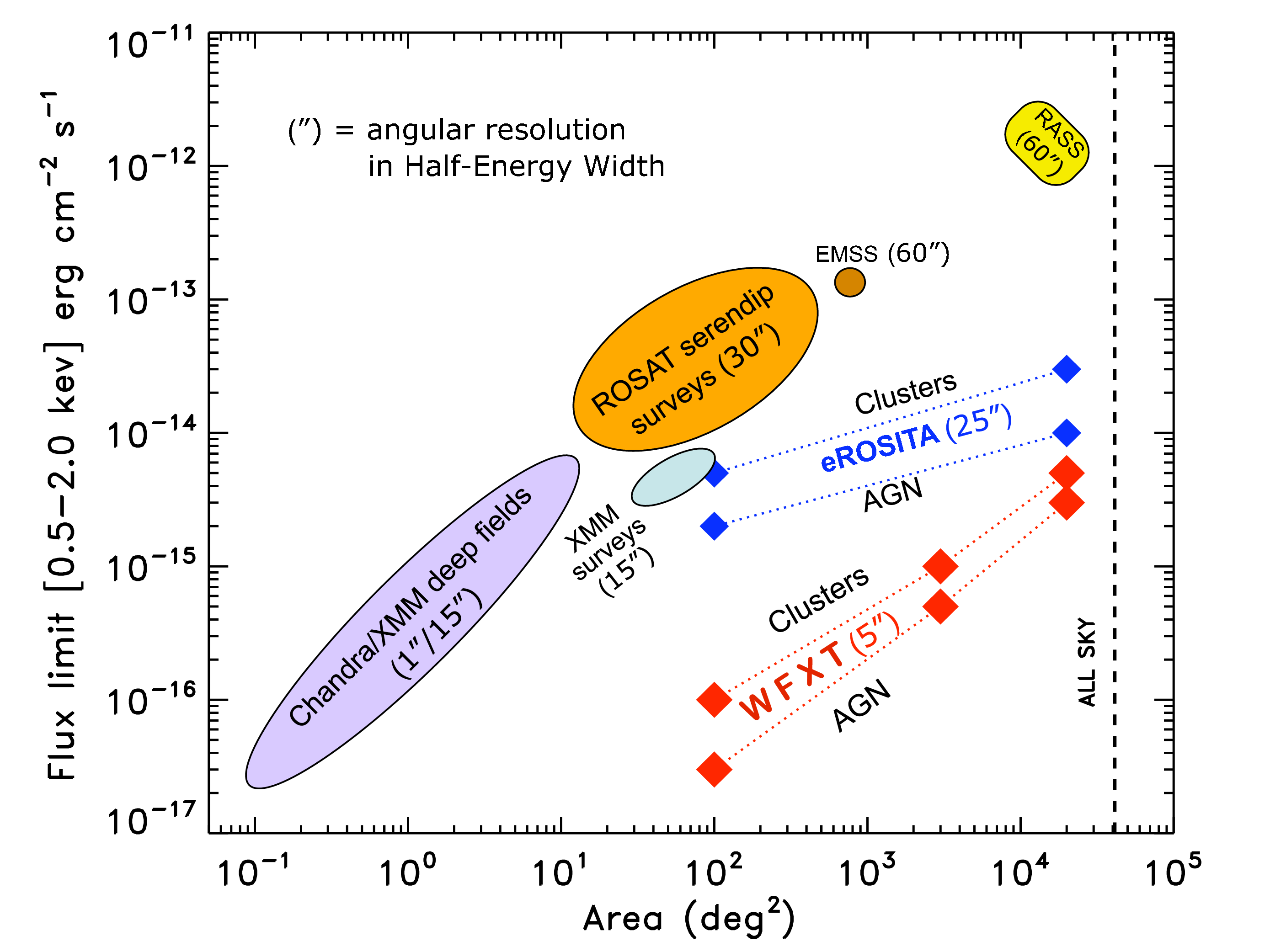}}
\raisebox{2.5cm}{\includegraphics[width=0.47\textwidth,clip=true]{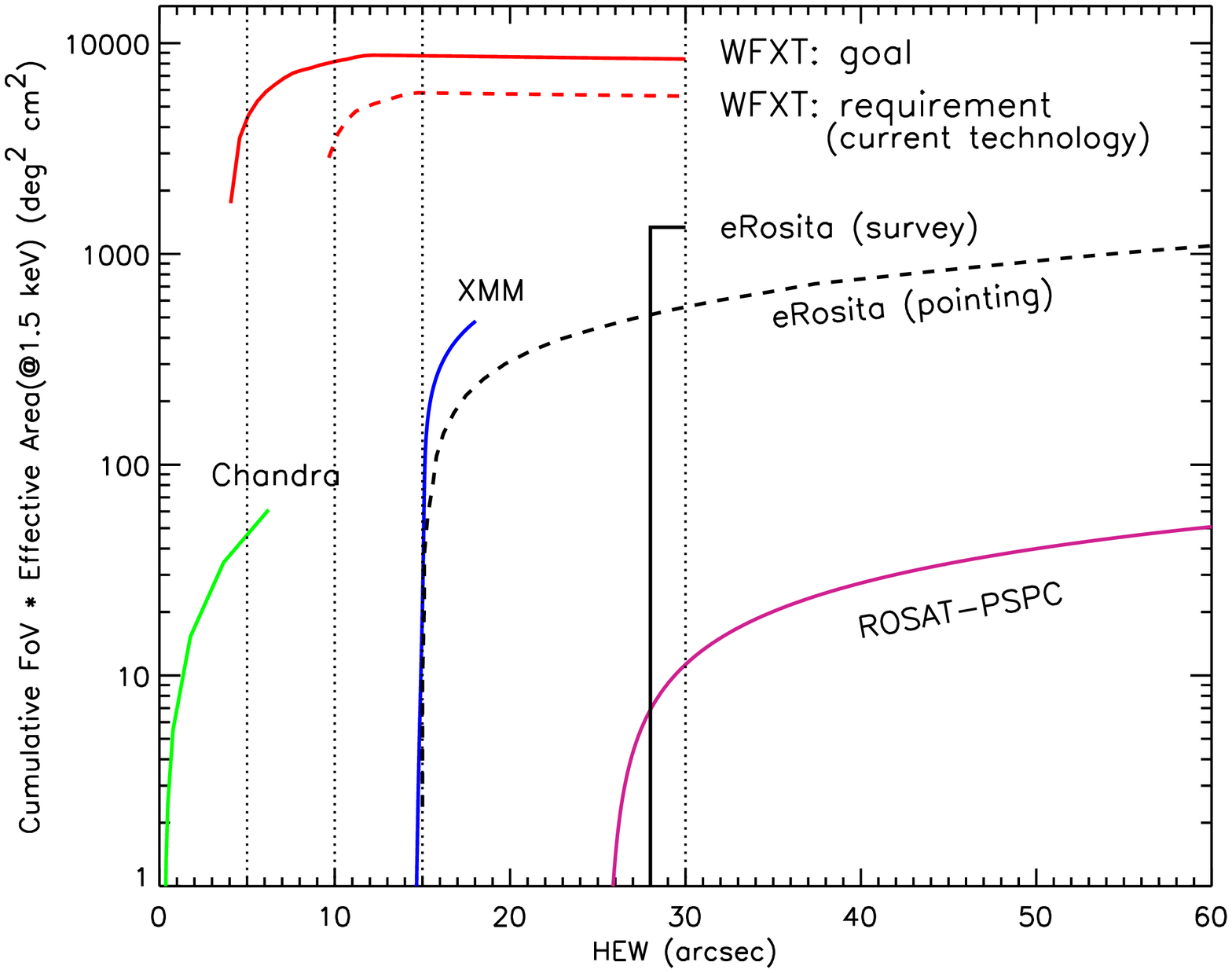}}
\vskip -2.2cm
\caption{ \footnotesize {\it Left}: Flux limits and sky
  coverage for past and planned X-ray surveys. The three WFXT surveys
  provide an unsurpassed combination of sensitivity and sky area.
  {\it Right}: WFXT cumulative Grasp $= \Omega \times \rm A$, as
  function of angular resolution (HEW).  WFXT's grasp is orders of
  magnitude greater than any other X-ray mission. Parameters for the
  planned mission {\it eROSITA} are taken from Cappelluti et al. in this
  volume. }
\end{center}
\label{f:surveys}
\end{figure*}

The high survey efficiency of WFXT, compared with other past or
planned X--ray missions, is obtained by using {\sl for the first time} a
wide-field optical design, first proposed by \citet*{BBG92}. By
adopting a polynomial shape of the X--ray mirros, WFXT's angular
resolution is optimized over the entire 1 deg$^2$ field of view, as
opposed to the classical Wolter-I optics whose angular resolution is
optimized mainly on-axis and degrades with the square of the off-axis
angle (see \citet{Conconi10}, \citet{Elsner10} and Pareschi et al. in
this volume). The resulting {\it Grasp}$=A\cdot\Omega_{\rm eff}$ of
the survey mission, i.e. the product of the telescope collecting area
and the effective field of view (FoV) at the desired angular
resolution, is significantly larger when compared to all other past or
proposed X-ray missions, making it an unprecedented survey instrument,
 able to carry out {\sl both wide and deep} surveys (see
Fig.\ref{f:surveys}/{\it left}). With such an enahanced discovery potential, WFXT
will provide optimum samples for both giant ground-based telescopes
for more sensitive, but narrow-field space facilities in the
optical-IR and X-ray. WFXT, however, is not only a path finder for
future missions, its large collecting area allows direct physical
characterization of a large fraction of sources (AGN and Clusters) via
X-ray spectroscopy with no need of follow-up observations. Synergy
with other missions further enhances its scientiﬁc potential and
breadth.  WFXT is conceived for the entire astronomical community.
Like the Sloan Digital Sky Survey
(SDSS\footnote{http://www.sdss.org/}), all WFXT data will become
public through a series of annual Data Releases that will constitute a
vast scientific legacy for decades.

\begin{table*}
\null\hskip 1.7cm
\parbox{10cm}{\caption{Description of the WFXT surveys$\, ^{(\star)}$}
\label{tab:surveys} }
\hbox{
\begin{tabular}{|l|l|l|l|}\hline
\multicolumn{1}{|c}{\raisebox{-2mm}{Quantity}} & \multicolumn{3}{|c|}{Survey}  \\ \cline{2-4}
  & \multicolumn{1}{c}{Deep} & \multicolumn{1}{|c}{Medium} &
  \multicolumn{1}{|c|}{Wide} \\ \hline \hline
$\Omega$ (deg$^2$) & 100 & 3000 & 20,000 \\ \hline
Exposure & 400 ksec & 13 ksec & 2 ksec \\ \hline
Total Time  $\, ^{(\star\star)}$ & 1.67 yr & 1.66 yr & 1.67 yr \\ \hline
$S_{\rm min}(0.5-2\,{\rm keV})$ point-like  & $3\times10^{-17}$ & $5\times10^{-16}$ & $3\times10^{-15}$ \\
\raisebox{1ex}{\ergscm \, at $5\sigma\, ^{(\star\star\star)}$} & \raisebox{1ex}{($1\times10^{-16}$)} &
   \raisebox{1ex}{($1\times10^{-15}$)} & \raisebox{1ex}{($5\times10^{-15}$)} \\ \hline
Total AGN detected & $5\times10^5$ & $4\times10^6$ & $1\times10^7$ \\ \hline
$S_{\rm min}(0.5-2\,{\rm keV})$ extended & $1\times10^{-16}$ & $1\times10^{-15}$ &$5\times10^{-15}$ \\
\raisebox{1ex}{\ergscm \, at $5\sigma$} & \raisebox{1ex}{($3\times10^{-16}$)} & \raisebox{1ex}{($2\times10^{-15}$)} & \raisebox{1ex}{($7\times10^{-15}$)} \\ \hline
Total Clusters/Groups & $3\times10^4$ & $2\times10^5$ & $3\times10^5$ \\ \hline
\end{tabular}\hskip 4mm
\parbox{3.5cm}{ \footnotesize 
 $^{(\star)}$ Values refer to goal performance parameters, those in
  parenthesis to minimal requirements of $A_{eff}=0.6\,{\rm m}^2,\,
  {\rm HEW}=10$\arcsec \\[2ex] 
 $\, ^{(\star\star)}$ Total observing time assumes 76\% observing efficiency 
\\[2ex] $^{(\star\star\star)}$ Flux limits in the hard 2--7 keV band are about 10 times higher}
}

\end{table*}

%
%

\section{Science goals and performance requirements}

To define the top level mission requirements, four major science cases
were identified and submitted as white papers to the Astro2010 Decadal
Survey of the National Academy of Sciences: {\sf 1)} Physics and
Evolution of Cluster of Galaxies \citep[Borgani et al. this
volume]{DS_Giacconi}; {\sf 2)} Growth and Evolution of Supermassive
Black Holes \citep[Gilli et al. this volume]{DS_Murray}; {\sf 3)}
Cosmology with Galaxy Clusters, \citep[Borgani et al. this
volume,][]{DS_Vik}; {\sf 4)} The very Local Universe \citep{DS_Ptak}.
Three extragalactic surveys, performed during five years of operation,
are required to fully meet the science goals described in these
papers: a {\it WIDE} survey covering most of the extragalactic sky
($\sim\!20,000$\ deg$^2$) at $\sim\!  500$ times the sensitivity, and
twenty times better angular resolution than the ROSAT All Sky Survey;
a {\it MEDIUM} survey mapping $\sim\! 3000$\ deg$^2$ to deep Chandra
and XMM sensitivity; and a {\it DEEP} survey probing $\sim\!  100$\
deg$^2$, ($\sim\!  1000$ times the area of the Chandra Deep Fields),
to the deepest Chandra sensitivity, with typical sampling timescales
of days to months.  Survey parameters are given in
Table~\ref{tab:surveys}. Flux limits and areas for the three surveys
are shown in Fig.\ref{f:surveys}/{\sl left}, along with those of
existing and planned X-ray surveys.

\begin{figure*}
\includegraphics[width=0.8\linewidth,angle=270]{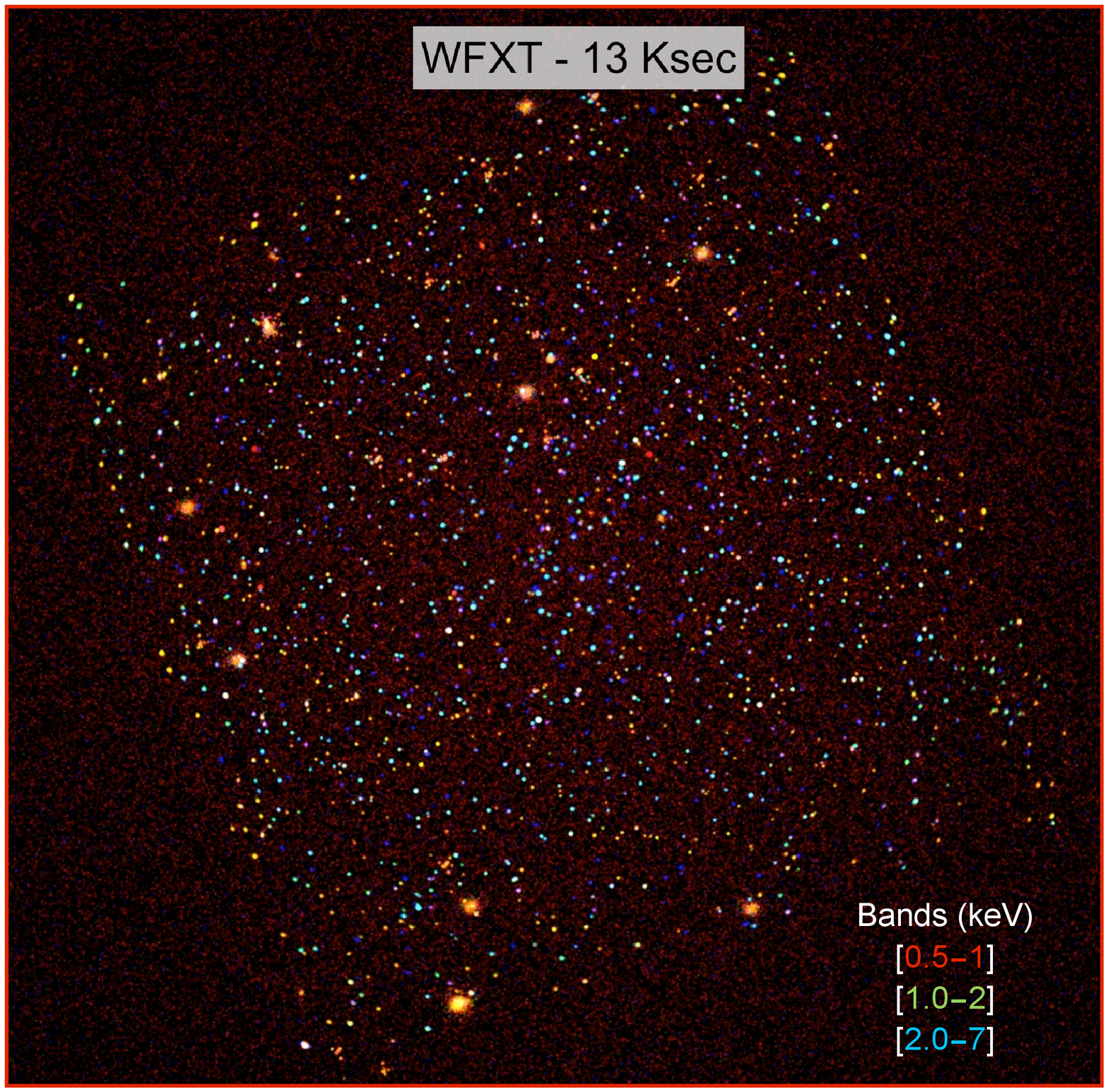} \\
\includegraphics[width=0.8\linewidth,angle=270]{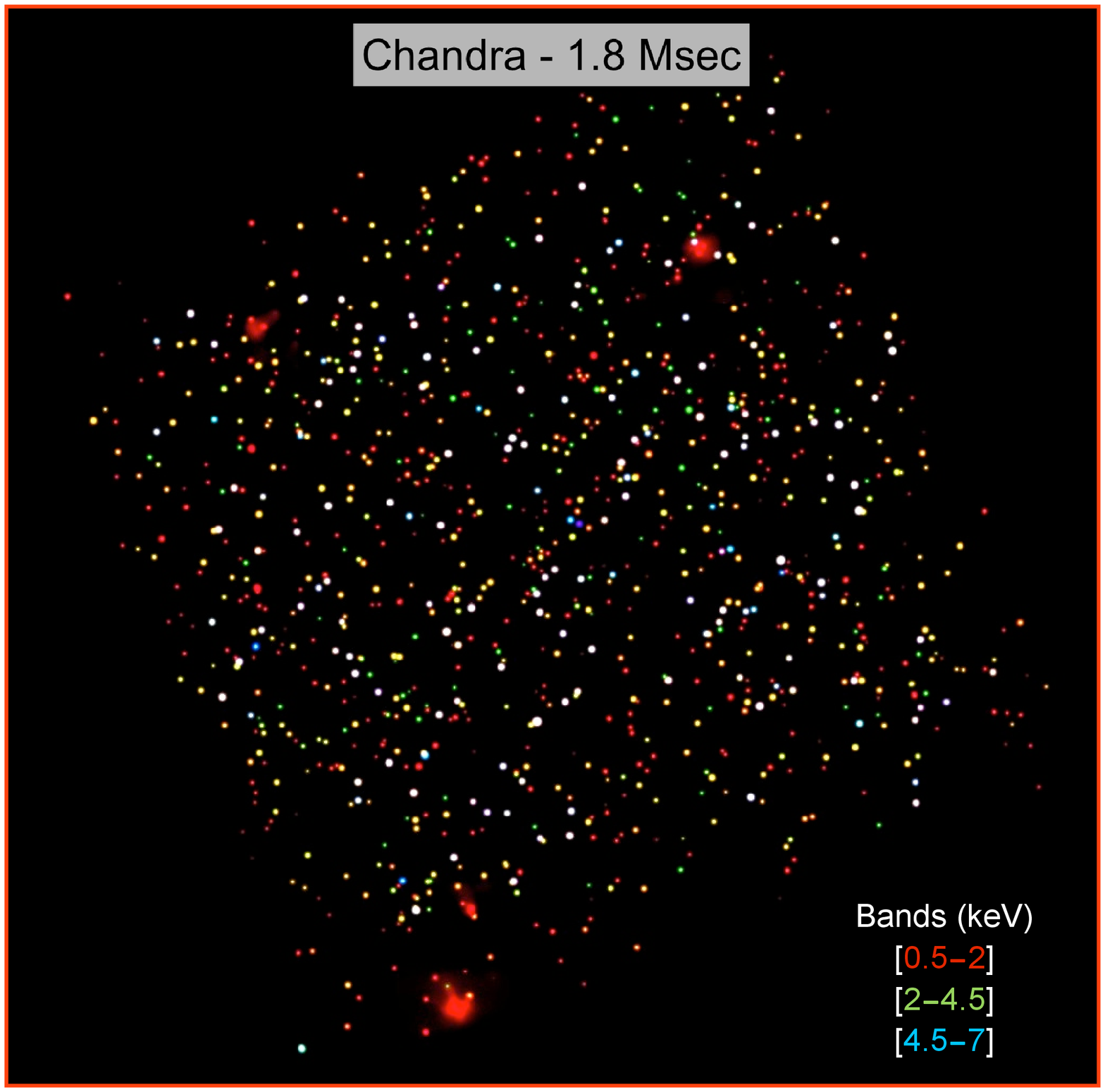}
\caption{\footnotesize Simulated WFXT image of the COSMOS field (top)
  observed with Chandra over 1 deg$^2$ (bottom) \citep{Elvis09}.  The
  flux limit of the two images is similar ($\sim\! 5\times
  10^{-16}\,\fun$ in 0.5-2 keV).  However the WFXT image (1 deg$^2$) is
  obtained with a single 13 ksec exposure (as part of the Medium
  survey), with an angular resolution (5\arcsec~HEW) close to
  Chandra's average ($\sim\! 3$\arcsec). The WFXT simulation was
  constructed from the Chandra COSMOS catalog \cite{Elvis09} with
  methods described in Tozzi et al. (this volume).  Bluer sources emit
  harder X-rays in the 0.5--7 keV band.}
\label{f:cosmos}
\end{figure*}

The capability of an X-ray observatory to carry out a survey, at a
given resolution, is given by the product of the {\it Grasp} defined
above and the time available for observation $T$. WFXT maximizes $A
\times \Omega_{\rm eff} \times T$ through its wide-field optics design
and a dedicated survey strategy which has an obvious advantage compared
to general facilities, such as \Chandra and {\it XMM-Newton}, which
have devoted $<10$\% of their time to surveys. The mirror design is no
more complex than a Wolter-I telescope, but yields a resolution of
5--10\arcsec\ over the entire 1 deg$^2$ field (Pareschi et al., this
volume).  To demonstrate the advantage of such a mission,
Fig.~\ref{f:surveys}/{\sl right} shows the cumulative field of view as
a function of angular resolution for five missions, derived from their
half-energy widths at different off-axis angles.  Even at 10\arcsec\
resolution, allowed by the wide-field design with current technology,
WFXT can survey a given area to a comparable flux limit in
$\sim\!\frac{1}{100}$ of the time that Chandra requires.  For example,
the simulation in Fig.\ref{f:cosmos} of a single 13 Ksec observation
(a {\it MEDIUM} survey tile) shows that WFXT would cover 1 deg$^2$ at
the sensitivity obtained by Chandra in a total observation of 1.8 Ms
of the COSMOS field \citep{Elvis09}. The wide-field X-ray optics and
mirror construction technology are also key to understand the clear
advantage with respect to the upcoming {\it eROSITA} survey mission
(Cappelluti et al., this volume) which provides a significant step
forward compared to the ROSAT All-Sky Survey.

\begin{figure*}
\hbox{
\parbox{5.5cm}{
\includegraphics[width=6.5cm,clip=true]{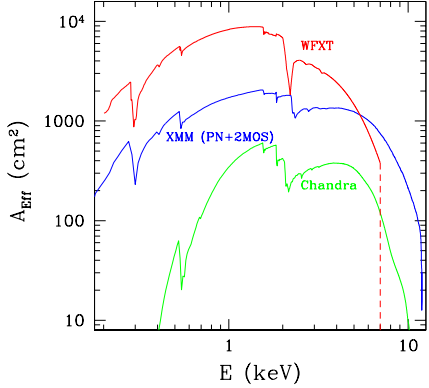} }
\hskip 3mm
\raisebox{0.1cm}{
\parbox{8cm}{~~~{\bf Table 2}: WFXT Mission Performance Requirements\\[1ex]
\begin{tabular}{|c|c|c|} \hline 
{\it Parameter} & {\it Requirement} & {\it Goal}\\ \hline \hline 
Area (1 keV) & $6,000\, cm^{2}$ & $10,000\,
cm^{2}$\\ \hline 
Area (4 keV) & $2,000\, cm^{2}$ & $3,000\, cm^{2}$\\ \hline 
Field of View & 1\deg\ diameter & 1.25\deg\ diameter\\ \hline 
Angular Resolution & $<10$\arcsec\ HEW & $\leq$5\arcsec\ HEW\\ \hline 
Energy Band & 0.2 - 5 keV & 0.1 - 7 keV\\ \hline 
Energy Resolution & \raisebox{0.3ex}{$\frac{E}{\Delta E}>10$} &  \raisebox{0.3ex}{$\frac{E}{\Delta E}>20$}\\ \hline
Time Resolution & $< 3$ seconds & $< 1$ second\\ \hline
Orbit & \multicolumn{2}{|c|}{550 km cir.,\, $<6^\circ$ incl.} \\ \hline
Mission Lifetime & \multicolumn{2}{|c|}{5 years} \\ \hline
\end{tabular} } }
}
\vskip -6mm
\caption{\footnotesize 
WFXT collecting effective area (goal) compared with \Chandra and \XMM.} 
\label{f:Aeff}

\end{figure*}
\stepcounter{table}

The angular resolution is a key parameter for the scientific success
of such a mission. A minimum requirement of 10\arcsec\ for the
half-energy width is dictated by the need to improve source
sensitivity, to discriminate point from extended emission, to minimize
source confusion and to allow an efficient identification of optical
counterparts. The latter is an essential process for the work-flow of
any science case. While a commonly used figure of merit for the {\sl
  discovery potential} of a survey mission is the {\it Grasp}, a more
appropriate figure of merit in X-ray surveys for the {\sl discovery
  speed}, i.e. the ability to {\sl discover and identify} sources
should scale as $A\Omega T\times{\rm HEW}^{-2}$, since the number of
possible counterparts scale with the area of the error circle. For
deep observations, no longer signal limited, the dependence on the
angular resolution will be even faster (up to ${\rm HEW}^{-3}$ in the
background limited regime).  Specifically, a 10\arcsec\ HEW yields a
1.5--2\arcsec positional accuracy thus easing the identification of
millions of sources, a daunting task with resolutions exceeding
10\arcsec. While 10\arcsec\ is feasible with present technology
(Pareschi et al. this volume), a goal of HEW~$\!\approx 5$\arcsec
approximately constant across the FoV has been chosen. This will {\it
  i)} further increase the sensitivity for point and extended sources
(e.g. groups), {\it ii)} enable AGN/cluster discernment at any
redshift, {\it iii)} enable confusion-free deep imaging, {\it iv)}
provide {\it Chandra}-like positional accuracy ($\lesssim 1$\arcsec)
with a source identification success rate of $>90$\% (Brusa et
al. this volume), {\it v)} resolve cool cores of $z\gtrsim 1$ clusters
\citep{Santos10} (essential for cosmological applications, see Borgani
et al. this volume), and {\it vi)} allow the detection of sharp
features (shocks, cold fronts, cavities) in the intra-cluster medium
(ICM).

\begin{table*}
\begin{center}
\caption{WFXT Science and Performance Requirements}\vskip -4mm
\label{tab:requirements}
\begin{tabular}{| p{0.25\textwidth} | p{0.7\textwidth} |}\hline
  \multicolumn{1}{|l|}{\it Performance Requirement} & \multicolumn{1}{|c|}{\it Observational/Science Requirements} \\
  \hline \hline 
  \null

  PSF HEW $<$10\arcsec (goal of 5\arcsec) across FoV & 
  {\em(i)} Sensitivity for point and extended sources 
  {\em(ii)} Minimize source confusion
  {\em(iii)} Discriminate extended sources from AGN at $z>1$
  {\em(iv)} Resolve cluster cool cores (50--100 kpc) at any redshift (10\arcsec\ corresponds to
  80 kpc at $z=1$) {\em(v)} efficient identification of optical counterparts  \\ \hline
  \null
  
  Large grasp = $\rm{FoV} \!\times\! A_{\rm Eff}$ 

  FoV $\ge 1\,{\rm deg}^2$

  $0.6 (0.2) < A_{\rm Eff}<\! 1 (0.3)\, {\rm m}^2$

  at 1(4) keV 
  & High survey efficiency and vast discovery space: {\em (a)} 
  to detect large numbers of sources thus measuring luminosity functions over a
  wide range of masses/luminosities and redshifts; {\em (b)} to guarantee high $S/N$ 
  spectra for significant subsets thus enabling physical characterization of clusters and AGN, including redshift measurements (with X-ray spectroscopy, without the need of follow-up observations); {\em (c)} to detect and characterize large number of variable and transient sources (AGN, GRB, SNe etc.) \\ \hline
  Low particle background achieved by low Earth orbit & 
  {\em (i)}  Improved limiting flux for point/extended sources; clusters out to large redshifts
  (proto-clusters at $z\sim\! 2$); {\em (ii)} Detect low surface brightness diffuse emission for nearby galaxies and clusters (out to and beyond the virial radius)
  \\ \hline
  Spectral resolution 

   ($10<E/\Delta E <20$)
& {\em (i)} Detect the Fe-K
  emission in clusters and AGN; {\em (ii)} Spectral characterization of
  clusters, AGN and galaxies \\ \hline
\end{tabular}
\end{center}
\end{table*}

The required effective area of the telescope at 1 kev is 0.6 m$^2$,
with a goal of up to 1 m$^2$, to still fit within the mass budget and
costs of a medium class mission with current or forseeen technology.
This large collecting area ultimately enables wide surveys at
unprecedented depth (Fig.\ref{f:surveys}) within the 5 year life time
of the mission, and thus allows very large volumes of the Universe to
be explored to large redshifts. In turn, this {\it i)} allows one to
trace the X-ray luminosity function of clusters and AGN (and
underlying mass functions of clusters and SMBHs) over a wide range of
masses and redshifts, and {\it ii)} enables physical characterization
of large samples of sources via their spectral analysis (see Borgani
et al. and Gilli et al. in this volume for predictions of expected
number of clusters and AGN and a detailed discussion of related science
cases).  As shown in Fig.\ref{f:Aeff}, it is also important to note
that the current (goal) design delivers an effective area  as
large as the one of {\it XMM} at 5 keV, this provides a significant benefit
when measuring temperatures and redshifts (using the Fe K line) from
the spectral fits of clusters as well as the detection of large
populations of obscured AGN (see Tozzi et al. in this volume).

\begin{figure*}
\begin{center}
\hbox {\hskip -4mm
\raisebox{-5mm}{\includegraphics[width=0.4\textwidth,clip=true, angle=270]{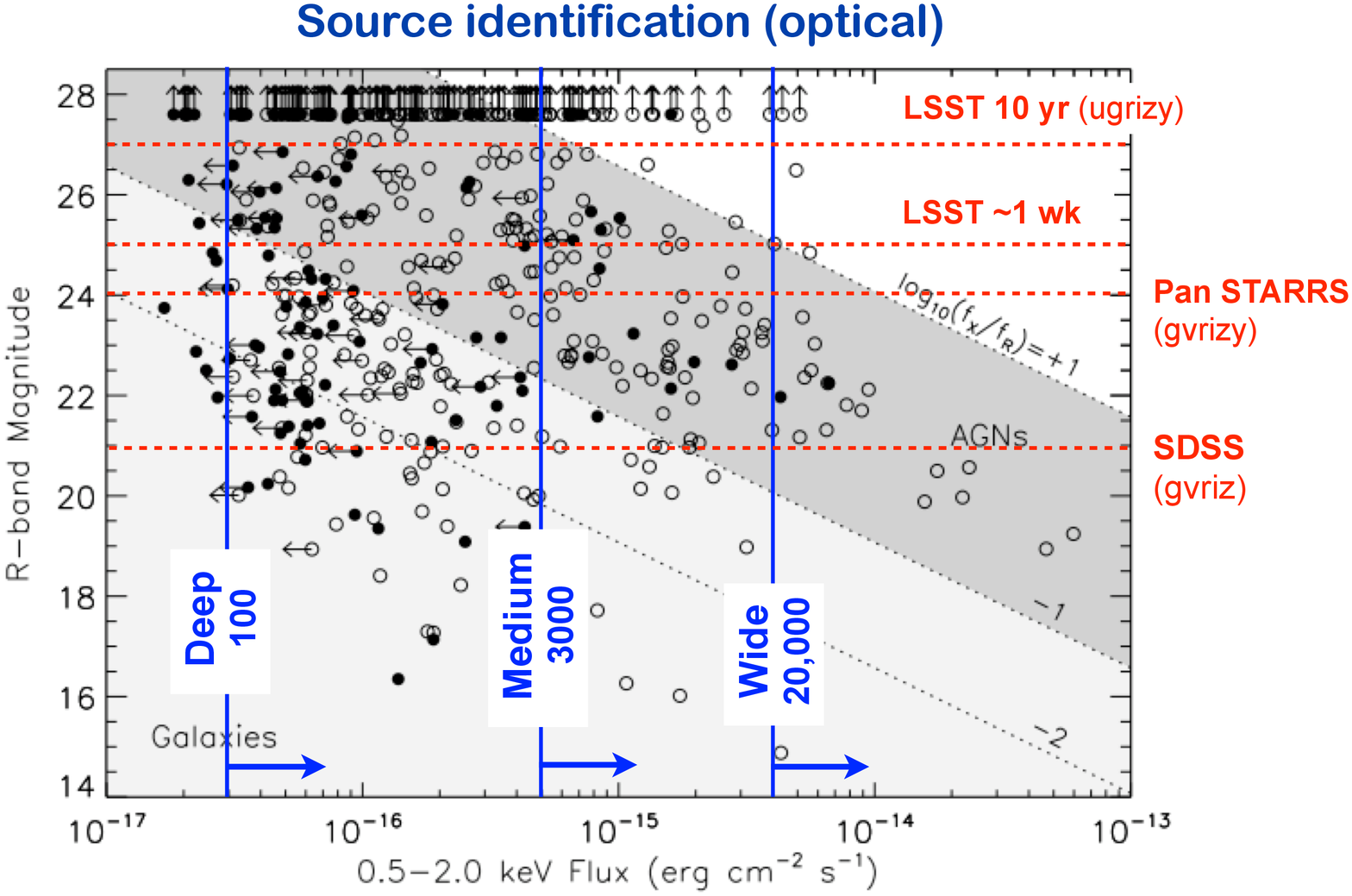}}
\hskip -0.9cm
\includegraphics[width=0.45\textwidth,clip=true, angle=270]{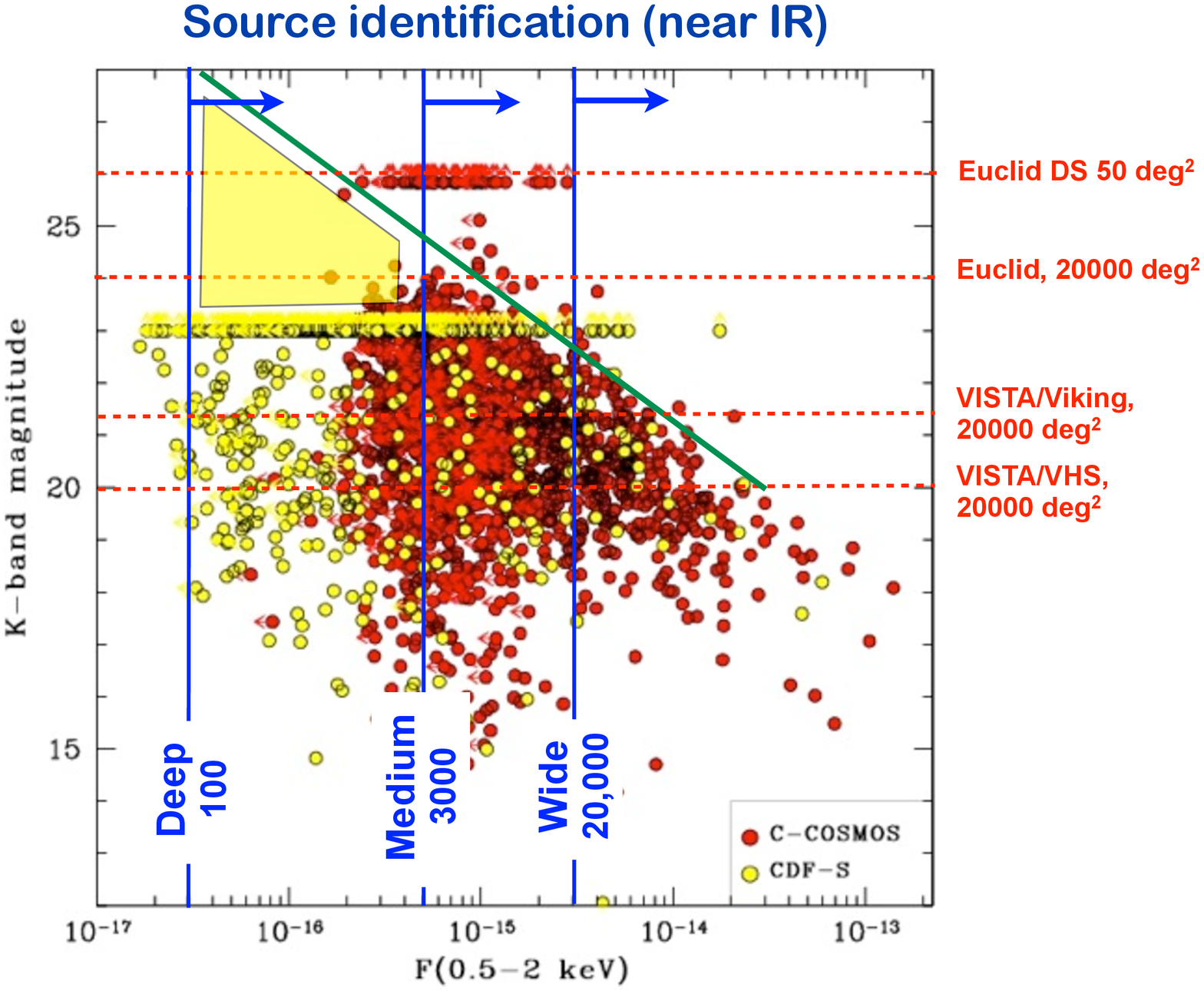}
}
\caption{ \footnotesize Source identification strategy for WFXT X-ray
  sources in the Optical and near-IR based on the distributions of
  flux ratios from \Chandra deep surveys in the R-band \citep{Luo08}
  and K$_{\rm AB}$-band (courtesy of C.Vignali).  Current and planned
  future wide-area surveys are indicated at different magnitude
  limits.}
\label{f:XrayIDs}
\end{center}
\end{figure*}

The $A\Omega$ and angular resolution combination also translates
in very interesting WFXT capabilities in the {\sl time domain},
enabling short temporal sampling observations and simultaneous
monitoring of large sky areas, which will allow one to detect and
study variable and transient X-ray populations of galactic and
extra-galactic sources (see Paolillo et al. this volume).

In addition, WFXT's low-Earth orbit reduces the particle background to
take full advantage of the instrument sensitivity and high-quality
PSF. This is important for the detection and spectral analysis of
low-surface brightness features such as distant groups, the most
distant clusters and the outskirts of nearby clusters (Ettori\&
Molendi, this volume).

We summarize in Table 2 the performance requirements and goals of the
mission, and in Table~\ref{tab:requirements} how these parameters are
connected to the observational and science requirements.

\section{Synergies with other surveys}

While we have emphasized how several science objectives can be
achieved with the WFXT data only using the X-ray spectral analysis of
significant sub-samples of sources with sufficient signal-to-noise,
there is no doubt that the synergy of WFXT with the next generation of
multi-wavelength deep wide-area surveys will greatly expand the
scientific grasp of the mission and will consolidate its vast legacy
value for decades.

The multi-wavelength properties of more than $10^7$ sources will be
available from the combination of current and future wide-area
surveys, such as
Pan-STARRS\footnote{http://pan-starrs.ifa.hawaii.edu/public/home.html}
and LSST\footnote{http://www.lsst.org} in the optical bands, and in
the near-IR surveys (VISTA, WISE, and JDEM/Euclid possibly by the end
of the decade), which will allow their identifcation and a measurement
of their photometric redshifts. Obtaining deep imaging in the near-IR
over the entire extra-galactic sky to identify the most obscured (and
most distant) AGN and distant clusters will remain the main challenge,
a task which only space-based surveys at $1-2\,\mu m$, such as those
proposed for the Euclid\footnote{http://sci.esa.int/euclid} or
JDEM\footnote{http://jdem.gsfc.nasa.gov/} missions, can perform. We
illustrate in Fig.\ref{f:XrayIDs} the identification strategies of AGN
for the three WFXT surveys, based on their optical/nearIR-to-X-ray flux
ratios measured in deep \Chandra surveys. The extension of the
multi-wavelength coverage to longer wavelengths, in the submm
(CCAT\footnote{http://www.submm.org/}) and radio with
SKA\footnote{http://www.skatelescope.org/} (see Padovani, this
volume), will complete the information on the spectral energy
distribution of different source populations with crucial implications
on our understanding of their physics and evolution.

For example, WFXT observations of thousands of clusters will provide
redshifts and detailed physical insights, such as temperature and
entropy profiles, metallicity of the ICM, mass density profiles of gas
and dark matter (DM). By combining this information with optical and
near-IR photometry of the cluster galaxies and with future high
sensitivity Sunyaev-Zeldovich surveys, one will be able to obtain a
comprehensive picture of the evolution of the baryons in their hot and
cold phases and how star formation and AGN activity affects the
physics of the ICM. By combining these data with lensing studies
carried out with ground (e.g. LSST) and space (e.g. Euclid)
observations, one will obtain detailed DM mass density profiles on a
range of redshifts and masses which can be compared with current
structure formation models thus setting strong constraints on the
 foundations of the \LCDM\ paradigm.

The spectroscopic follow-up study of large subsamples of WFXT sources
will remain a serious challenge. Dedicated wide-field,
high-multiplexing spectrographs on 8m class telescopes currently under
study will be suitable for a wide range of science cases. The
systematic spectroscopic identification work in large ($\sim\! 10^4$
deg$^2$) survey areas can partly be carried out with near-IR slitless
surveys, which are part of the Euclid/JDEM mission concepts. A more effective
approach to the study of a large variety of sources over a wide redshift
range would require wide-area {\sl slit}-spectroscopy from space, such
as the one proposed for the SPACE\citep{SPACE} mission.

Moreover, in combination with more sensitive, narrow-field
observatories, WFXT will be an outstanding source of interesting high
redshift clusters and AGN for follow-up studies with JWST (or its
successors), ALMA, the next generation of giant (30-40m) ground-based
telescopes, and X-ray observatories (i.e., IXO and Gen-X).

\section{Conclusions}

In this article, we have illustrated the concept of the WFXT mission
and how the mission requirements flow from the main science drivers
and observational requiremens. We refer the reader to all the
contributions in this volume for a detailed discussion of several
science cases, which range from the formation and evolution of SMBHs
and clusters to stellar populations and compact objects in the Galaxy;
from Cosmology to the physics of clusters and AGN, including the study
of early-type and star-forming galaxies. This collection of science
cases is by no means complete, but can be considered the basis on
which the scientific potential of WFXT can be further explored as the
technological development continues. It is important to emphasize that
the gain margin of WFXT compared to previous or planned X-ray
missions in conducting surveys is so large that its scientific impact
would remain very strong even if cost or technological challenges will
drive a redefinition of its performance parameters. In the suite of
requirements under study, as explained above, the angular resolution
remains the one parameter on which is very difficult to compromise, as
\Chandra observations have unambigously and definitively taught us.

When examining the range of wide-area high-sensitivity surveys being
planned for this decade, in the optical, IR, submm and radio regimes,
WFXT stands out as the only one which will be able to match these
surveys in coverage, sensitivity and angular resolution at soft X-ray
wavelengths. Coordinated surveys from X-ray to radio wavelengths,
which have been carried out in small areas of the sky ($\lesssim 1$
deg$^2$) over the last decade, have definitively established the
crucial value of the multi-wavelength approach in astrophysics which
has fueled phenomenal progress in many areas. X-ray observations have
been key to such a progress as they have the unique ability to probe
phenomena and unveil sources powered by {\sl gravity}.  On the other
hand, systematic wide-area surveys have demonstrated that they are
able to produce major discoveries and address fundamental questions,
as also underscored by their high level of high-impact publications
\citep{Madrid09}.  We therefore argue that the lack of a mission
like WFXT in the suite of future multi-wavelength wide-area surveys
will ultimately limit their scientific potential.

\begin{acknowledgements}
  We are grateful to Colin Norman and the entire WFXT Team for intense
  discussions over the years which have led to the current mission
  concept and development.

\end{acknowledgements}

\bibliographystyle{aa}

\end{document}